\documentclass[11pt,a4paper]{article}
\usepackage{jheppub}

\usepackage[utf8]{inputenc}
\usepackage{amsmath,amssymb}
\usepackage{slashed}

\newcommand{\ii}{\mathrm{i}}
\newcommand{\bM}{\partial\mathcal{M}}

\title{Gravitational parity anomaly with and without boundaries}

\author[a,b,c]{Maxim Kurkov}
\author[a,d]{and Dmitri Vassilevich}

\affiliation[a]{CMCC, Universidade Federal do ABC\\
Avenida dos Estados 5001, CEP 09210-580, Santo Andr\'e, SP, Brazil}
\affiliation[b]{Dipartimento di Fisica “E. Pancini”, Universita di Napoli “Federico II”,\\ Complesso Univ. Monte S. Angelo, Via Cintia, I-80126 Napoli, Italy}
\affiliation[c]{INFN - Sezione di Napoli,  Via Cintia, 80126 Napoli, Italy}
\affiliation[d]{Department of Physics, Tomsk State University,\\ 36 Lenin Ave, 634050 Tomsk, Russia}
\emailAdd{max.kurkov@gmail.com}
\emailAdd{dvassil@gmail.com}

\abstract{In this paper we consider gravitational parity anomaly in three and four dimensions. We start with a re-computation of this anomaly on a 3D manifold without boundaries and with a critical comparison of our results to the previous calculations. Then we compute the anomaly on 4D manifolds with boundaries with local bag boundary conditions. We find, that gravitational parity anomaly is localized on the boundary and contains a gravitational Chern-Simons terms together with a term depending of the extrinsic curvature. We also discuss the main properties of the anomaly, as the conformal invariance, relations between 3D and 4D anomalies, etc.}

\begin{document}
\maketitle

\section{Introduction}\label{sec:Int}
It is known for quite a long time \cite{Niemi:1983rq,Redlich:1983dv} that in a quantum theory of a Dirac fermion interacting with an external gauge filed in odd dimensions one cannot simultaneously preserve parity and gauge invariance. Thus gauge invariant quantum theories inevitably possess the so-called parity anomaly which manifests itself through a Chern-Simons term  in the one-loop effective action. Alvarez-Gaume \emph{et al} \cite{AlvarezGaume:1984nf} extended these arguments to odd-dimensional fermions interacting with gravitational backgrounds and related parity anomaly to the spectral asymmetry of Dirac operator. 

In a recent paper \cite{Kurkov:2017cdz} we demonstrated that also in four dimensions Dirac fermions subject to local boundary conditions interacting with an abelian gauge field exhibit the parity anomaly which leads a Chern-Simons term on the boundary. The purpose of this paper is to extend this result to gravitational backgrounds. Our work was motivated by the results of \cite{Witten:2015aba} where interesting relations between bulk and boundary anomalies following from the Atiyah-Patodi-Singer (APS) index theorem were established. In another paper \cite{Muller:2017jzu}, extensions of the APS index theorem to manifolds with corners were addressed. Domain wall fermions in this context were considered in \cite{Mulligan:2013he,Fukaya:2017tsq}. From the physical point of view, parity anomaly is related to the Hall conductivity on the surface of Topological Insulators, which is currently under discussion \cite{Konig:2014ema}.  Other boundary anomalies also received considerable attention recently, see for instance \cite{Herzog:2015ioa,Fursaev:2015wpa,Solodukhin:2015eca,Jensen:2017eof}. We would also like to mention that the parity anomaly plays a role in the context of the boson-fermion duality, see  \cite{Ferreiros:2018ohl} and references therein.

In this paper we use the $\zeta$ function regularization and the heat kernel methods. The mathematical background may be found in \cite{GilkeyNew,Kirsten:2001wz,Vassilevich:2003xt}. The parity anomaly is understood as the $\eta(0)$ invariant of the Dirac operator. General expressions needed to compute the anomaly are collected in Section \ref{sec:heat}. 

We start actual computations in Section \ref{sec:3D} where we reconsider the parity anomaly on 3D manifolds without boundaries. We reconfirm that the anomaly in given by gravitational Chern-Simons action, first introduced in \cite{Deser:1981wh,Deser:1982vy}. (See \cite{Zanelli:2012px} for a review on Chern-Simons forms in gravity.) There have been contradicting results in the literature regarding the coefficient in front of Chern-Simons term. We comment on these contradictions at the end of Section \ref{sec:3D}.

The main material of this paper is contained in Section \ref{sec:4D} where we study 4D manifolds with boundaries. In contrast to the paper \cite{Witten:2015aba} which used the non-local APS boundary conditions, our boundary conditions are local bag-type. The latter conditions are much easier to realize in physics, but they do not respect chirality. We find that parity anomaly contains apart from the Chern-Simons term also a term depending on extrinsic curvature of the boundary. Finally, we check conformal invariance of the anomaly and discuss relations to 3D anomalies and topological densities.

\section{Parity anomaly through the heat kernel expansion}\label{sec:heat}
In this Section we shall express the parity anomaly through the spectral $\eta$ function and then compute the variation of the latter in terms of the heat kernel expansion. This method was used in \cite{Fursaev:2011zz,Vassilevich:2007gt,Kurkov:2017cdz}. It is based on the earlier papers \cite{AlvarezGaume:1984nf,Deser:1997nv,Deser:1997gp}. We shall follow \cite{Fursaev:2011zz} almost literally, so that just a short overview will be enough.

The $\zeta$ function of Dirac operator $\slashed{D}$ depends on a complex spectral parameter $s$ and is defined through summation over the eigenvalues $\lambda$
\begin{equation}
\zeta(s,\slashed{D})=\sum_{\lambda>0}\lambda^{-s} +e^{\ii \pi s}\sum_{\lambda<0}(-\lambda)^{-s}\,.
\label{zetaD}
\end{equation}
The sums above are convergent for $\Re s>n$, where $n={\rm dim}\, \mathcal{M}$. From this region, the zeta function can be continued as a meromorphic function to the whole complex plane.
In terms of this function, the $\zeta$-regularized effective action reads
\begin{equation}
W_s=-\ln \det (\slashed{D})_s=\mu^s\Gamma(s)\zeta(s,\slashed{D})\,,\label{Ws}
\end{equation}
where we introduced a parameter $\mu$ of mass dimension 1 to make $W_s$ dimensionless. The physical limit (lifted regularization) is $s\to 0$.

Let us separate the parts of $\zeta$ function that are even and odd with respect to the reflection $\slashed{D}\to - \slashed{D}$,
\begin{eqnarray}
\zeta(s,\slashed{D})&=&\zeta(s,\slashed{D})_{\rm even}+\zeta(s,\slashed{D})_{\rm odd}\nonumber\\
\zeta(s,\slashed{D})_{\rm even}&=& \tfrac 12 \bigl(\zeta(s,\slashed{D})+\zeta(s,-\slashed{D})\bigr),
\label{zeven}\\
\zeta(s,\slashed{D})_{\rm odd}&=& \tfrac 12 \bigl(\zeta(s,\slashed{D})-\zeta(s,-\slashed{D})\bigr).
\label{zodd}
\end{eqnarray}
With the help of spectral $\eta$ function
\begin{equation}
\eta(s,\slashed{D})=\sum_{\lambda>0}\lambda^{-s}-\sum_{\lambda<0}(-\lambda)^{-s} \label{etas}
\end{equation}
the odd part can be rewritten as
\begin{equation}
\zeta(s,\slashed{D})_{\rm odd}=\tfrac 12\bigl( 1-e^{\ii \pi s}\bigr) \eta(s,\slashed{D}) .\label{zetaeta}
\end{equation}
Corresponding part of the effective action is finite at $s\to 0$
\begin{equation}
W^{\rm odd}=W^{\rm odd}_{s=0}=\frac {\ii \pi}2 \eta(0,\slashed{D}) \label{Wodd}
\end{equation}
and is identified with the parity anomaly. This is a known formula, see \cite{AlvarezGaume:1984nf}.

To evaluate the anomaly, we shall use the following integral representation for $\eta$ function
\begin{equation}
\eta(s,\slashed{D})=\frac 2{\Gamma\left( \frac {s+1}2 \right)} \int_0^\infty d\tau\, \tau^s \mathrm{Tr}
\left(\slashed{D}e^{-\tau^2 \slashed{D}^2} \right)\,.\label{etaint}
\end{equation}
Let us consider the variation of $\eta$ under a small variation of the Dirac operator, $\slashed{D}\to\slashed{D}+\delta\slashed{D}$,
\begin{equation}
\delta\eta(s,\slashed{D})=\frac 2{\Gamma\left( \frac {s+1}2 \right)} \int_0^\infty d\tau\, \tau^s \partial_\tau \mathrm{Tr}
\left((\delta \slashed{D})\tau e^{-\tau^2 \slashed{D}^2} \right)\,.\label{etavar}
\end{equation}
After taking $s=0$ and computing the integral, we arrive at
\begin{equation}
\delta\eta(0,\slashed{D})=-\frac 2{\sqrt{\pi}} \lim_{t\to +0} \mathrm{Tr}
\left((\delta \slashed{D})t^{1/2} e^{-t \slashed{D}^2} \right)\,.\label{deta}
\end{equation}

The right hand side of (\ref{deta}) will be evaluated with the help of the heat kernel expansion.
The variation $\delta\slashed{D}$ is, in general, a first-order differential operator. For any first order operator $Q$ and any Laplace type operator $L$ with local boundary conditions there is a full asymptotic expansion as $t\to +0$, see \cite{GilkeyNew},
\begin{equation}
{\rm Tr}\, \left( Qe^{-tL} \right) \simeq \sum_{k=-1}^\infty t^{\frac{k-n}2} a_k(Q,L)\,.\label{hkex}
\end{equation}
By collecting everything together, we obtain
\begin{equation}
\delta W^{\rm odd}=-\ii \sqrt{\pi}\, a_{n-1}(\delta\slashed{D},\slashed{D}^2)\,.\label{dWodd}
\end{equation}

General Laplace type operator may be written in the form
\begin{equation}
L=-(\nabla\cdot\nabla +E)\,,\label{Lap}
\end{equation}
where $\nabla=\partial+\omega$ is a covariant derivative, $E$ is a zeroth order part, dot means the contraction with a Riemannian metric. The expression $\nabla\cdot\nabla$ contains also the Christoffel connection.

Usually, the heat kernel coefficients are computed with $Q$ being the identity operator or a smooth function without any matrix structure. However, the desired expressions $a_k(Q,L)$ may be recovered with the help of a simple variational procedure, see \cite{Branson:1997ze}. Let us write 
\begin{equation}
Q=q_1\cdot \nabla + Q_0, \label{Qq1}
\end{equation}
where $Q_0$ is a zeroth order operator, and define $L(y,z)$ as the operator $L$ with the replacements $E\to E+yQ_0$, $\omega\to\omega+\tfrac 12 z q_1$. Then $L=L(0,0)$ and
\begin{eqnarray}
&& \frac{\partial}{\partial y}\vert_{y=z=0} {\rm Tr}\, \left( e^{-tL(y,z)}\right)=
t\, {\rm Tr}\, \left( Q_0 e^{-tL}\right)\,,\label{var1}\\
 && \frac{\partial}{\partial z}\vert_{y=z=0} {\rm Tr}\, \left( e^{-tL(y,z)}\right)=
t\, {\rm Tr}\, \left( \left(q_1\cdot \nabla +\tfrac 12 (\nabla \cdot q_1)\right) e^{-tL}\right)\,. \label{var2}
\end{eqnarray}
Expanding these equations in the asymptotic series we get
\begin{eqnarray}
&& \frac{\partial }{\partial y}\vert_{y=z=0}\, a_{k+2}(1,L(y,z))=a_k(Q_0,L)\,,\label{vara1}\\
&& \frac{\partial }{\partial z}\vert_{y=z=0}\, a_{k+2}(1,L(y,z))=a_k\left(q_1\cdot \nabla +\tfrac 12 (\nabla \cdot q_1) ,L\right)\,.\label{vara2}
\end{eqnarray}
Actually, all heat kernel coefficients weighted with a zeroth order operator that will be needed in this paper can be found in the literature. We presented Eq.\ (\ref{vara1}) for the sake of completeness. 

\section{Three dimensions, no boundaries}\label{sec:3D}
In this Section, $\mathcal{M}$ is a 3-dimensional manifold without boundaries. We use $\mu,\nu\rho,...$ as world indices on $\mathcal{M}$ while $a,b,c,...$ denote flat (tangential) indices. The Dirac operator reads
\begin{equation}
\slashed{D}=\ii \gamma^a e_a^\mu\nabla_\mu, \qquad \nabla_\mu = \partial_\mu+\tfrac 18 \sigma_{\mu ab}[\gamma^a,\gamma^b]\,,\label{Dop3}
\end{equation}
where $e_a^\mu$ is a dreibein, $\sigma_{\mu ab}$ is a spin-connection. We take $\gamma^a=\sigma^a$ with $\sigma^a$ being Pauli matrices. Therefore, $\gamma^a\gamma^b+\gamma^b\gamma^a=2\delta^{ab}$ and
\begin{equation}
\mathrm{tr}\, (\gamma^a\gamma^b\gamma^c)=2\ii \epsilon^{abc},\qquad \epsilon^{123}=1.
\end{equation}

Let us consider the variations of dreibein 
\begin{equation}
\delta e_{\mu a}=u_{ab}e^b_\mu \label{vare}
\end{equation}
parametrized by an infinitesimal parameter $u_{ab}$. A skew-symmetric $u_{ab}$ corresponds to a local frame rotation and thus belongs to the gauge symmetries of the theory. Therefore, for our purposes it is sufficient to consider a symmetric $u_{ab}=u_{ba}$. Then $\delta g_{\mu\nu} = 2e_\mu^a e_\nu^b u_{ab}=2u_{\mu\nu}$ and
\begin{equation}
\delta \sigma_{\mu ab}=-u_{b\mu;a} + u_{a\mu;b}\,,\label{delsi}
\end{equation}
where semicolon is used to denote covariant derivatives. For example, $\nabla_\mu \nabla_\nu u_{ab}\equiv u_{ab;\nu \mu}$. The corresponding variation of $\slashed{D}$ reads
\begin{equation}
\delta \slashed{D}=-\ii \gamma^a u_a^{\ \mu}\nabla_\mu +\tfrac \ii 4 u_{\mu a;b}[\gamma^a,\gamma^b]\,.
\label{varD}
\end{equation}
Thus $\delta\slashed{D}=Q$ with
\begin{equation}
q_1^\mu =-\ii \gamma^a u_a^{\ \mu}, \qquad Q_0=\tfrac \ii 4 u_{\mu a;b}[\gamma^a,\gamma^b] \label{q1Q0}
\end{equation}
(cf Eq.\ (\ref{Qq1})).
With the help of (\ref{a2}) one easily checks that 
\begin{equation}
a_2\bigl( Q_0,\slashed{D}^2\bigr)=\tfrac 12 a_2\bigl((\nabla \cdot q_1),\slashed{D}^2\bigr)=0\,.
\end{equation}
Therefore, the variation of $W^{\rm odd}$ is given just by the heat kernel coefficient $a_2(q_1\cdot \nabla,\slashed{D}^2)$.
To compute this coefficient we proceed as prescribed by Eq.\ (\ref{vara2}). We shift the connection $\omega$, eq.\ (\ref{omE}),
\begin{equation}
\omega_\mu \to \omega_\mu(z)=\omega_\mu -\frac {\ii z} 2 \gamma^a u_{a\mu}\,, \label{omz}
\end{equation}
so that for the corresponding field strength $\Omega_{\mu\nu}=[\nabla_\mu,\nabla_\nu]$, see (\ref{Om}), we have
\begin{equation}
\Omega_{\mu\nu}(z)=\Omega_{\mu\nu}+\ii z\gamma^a \mathcal{Q}_{a\mu\nu} +\mathcal{O}(z^2), \qquad \mathcal{Q}_{a\mu\nu}\equiv \tfrac 12 \bigl( u_{a\mu;\nu}-u_{a\nu;\mu}\bigr).
\label{Omz}
\end{equation}
Just a single term $\Omega^2$ in (\ref{a4}) contributes to $(\partial_z \vert_{z=0} a_4(1,L(0,z))$, so that
\begin{equation}
\delta W^{\rm odd}=-\ii \sqrt{\pi}a_2(q_1\cdot \nabla,\slashed{D}^2)=-\frac {\ii}{96\pi} \int d^3x \sqrt{g} \epsilon^{abc} u_{a\mu}R^{\mu\nu}_{\ \ \ bc;\nu}\,. \label{delW3}
\end{equation}
With the help of the variational identity
\begin{equation}
\delta \int d^3x \sqrt{g} \epsilon^{\mu\nu\rho}\left( \Gamma_{\mu\kappa}^\lambda \partial_\nu \Gamma_{\rho\lambda}^{\kappa}+\tfrac 23 \Gamma_{\mu\kappa}^\lambda \Gamma_{\nu\sigma}^\kappa \Gamma_{\rho\lambda}^\sigma\right) =2  \int d^3x \sqrt{g} \epsilon^{abc} u_{a\mu}R^{\mu\nu}_{\ \ \ bc;\nu}\label{varCS}
\end{equation}
we obtain
\begin{equation}
 W^{\rm odd}=-\frac {\ii}{192\pi} \int d^3x \sqrt{g} \epsilon^{\mu\nu\rho}\left( \Gamma_{\mu\kappa}^\lambda \partial_\nu \Gamma_{\rho\lambda}^{\kappa}+\tfrac 23 \Gamma_{\mu\kappa}^\lambda \Gamma_{\nu\sigma}^\kappa \Gamma_{\rho\lambda}^\sigma\right)\,. \label{W3}
\end{equation}
This is the main result of this section.

As a consistency check one has to verify that the coefficients $a_0$ and $a_1$ do not contribute to (\ref{deta}) so that the right hand side of (\ref{deta}) remains finite at $t\to 0$. We leave this to the reader as an exercise.

The gravitational parity anomaly in 3D was computed in \cite{Goni:1986cw,Vuorio:1986ju,vanderBij:1986vn,Ojima:1988av}.  We agree with the results of \cite{Goni:1986cw,Vuorio:1986ju,vanderBij:1986vn} (note Erratum in Ref. \cite{Vuorio:1986ju}) but disagree with Ojima \cite{Ojima:1988av}. The paper by Ojima computed a fraction of two determinants, $\det (\slashed{D}-\ii m)/\det(\slashed{D}_{\sigma=0}-\ii m)$, see \cite[Eq.\ (3.3)]{Ojima:1988av}, where $\slashed{D}_{\sigma=0}$ means the Dirac operator in curved space but without the spin-connection term. Obviously, this is quite different to what we did, since we considered the variation of geometry in both terms of the Dirac operator. Moreover, since the operator $\slashed{D}_{\sigma=0}$ is not hermitian, $\det \slashed{D}_{\sigma=0}$ does not correspond to the path integral in any consistent theory and the parity odd part of the fraction of determinants cannot be interpreted through the $\eta$ function. Quite naturally, our results differ from that of \cite{Ojima:1988av}.

\section{Four dimensional manifold with boundaries}\label{sec:4D}
\subsection{Boundary value problem}\label{sec:bvp}
In this section, $\mathcal{M}$ is a four-dimensional Riemannian manifold with a smooth boundary $\bM=\bigcup_\alpha \bM_\alpha$, where $\bM_\alpha$ denotes different connected components. Let $n$ be the inward pointing unit normal to the boundary. Let $\gamma^n=n_\mu \gamma^\mu$. We impose local bag boundary conditions that read in the Euclidean signature
\begin{equation}
\Pi_-\psi \vert_{\bM}=0,\qquad \Pi_-=\tfrac 12 (1-\ii \varepsilon_\alpha \gamma^5\gamma^n)\,.
\label{1stbc}
\end{equation}
$\varepsilon_\alpha=\pm 1$ is constant on each of the components $\bM_\alpha$, but may vary on $\bM$. The chirality matrix is defined as
\begin{equation}
\gamma^5=\frac 1{4!} \epsilon^{\mu\nu\rho\sigma}\gamma_\mu \gamma_\nu \gamma_\rho \gamma_\sigma \,,
\label{gam5}
\end{equation}
where $\epsilon^{\mu\nu\rho\sigma}$ is the Levi-Civita tensor. With this sign convention
\begin{equation}
\mathrm{tr}\, \bigl( \gamma^5 \gamma_\mu \gamma_\nu \gamma_\rho \gamma_\sigma \bigr)
=4 \epsilon^{\mu\nu\rho\sigma}\,.\label{trg5}
\end{equation}

$\Pi_-$ is a projector, $\Pi_-^2=\Pi_-$. For the future use, we define also a complementary projector $\Pi_+:=1-\Pi_-$. These boundary conditions appeared for the first time in the context of bag model of hadrons \cite{Chodos:1974je,Chodos:1974pn} and were rediscovered in mathematical literature in \cite{Branson:1992}, see also \cite{Luckock:1990xr}. For these boundary conditions the current of fermions through the boundary vanishes and thus $\slashed{D}$ is symmetric, see (\ref{B3}). However, to make $\slashed{D}$ selfadjoint one has to impose a second boundary condition 
\begin{equation}
\Pi_- \slashed{D}\psi \vert_{\bM}=0 ,\label{2ndbc}
\end{equation}
which can be written also as
\begin{equation}
(\nabla_n + S)\Pi_+ \psi\vert_{bM}=0,\qquad S=-\tfrac 12 \Pi_+ K \label{2bc2} 
\end{equation}
with $K$ being the trace of extrinsic curvature on $\bM$. We also define
\begin{equation}
\chi\equiv \Pi_+-\Pi_- \,. \label{chi}
\end{equation}

Note, that since the boundary projector $\Pi_-$ (\ref{1stbc}) contains the chirality matrix $\gamma^5$, bag boundary conditions cannot be defined on odd-dimensional manifolds of Euclidean signature. 

To simplify computations we shall use the Gaussian normal coordinates near $\bM$ with the line element
\begin{equation}
(ds)^2=h_{ij}dx^idx^j +(dx^n)^2,\label{ds2}
\end{equation}
so that $x^n$ will be the normal geodesic coordinate, while $x^i,x^j,x^k,\dots$ will be coordinates on $\bM$. A local orthonormal frame $e_A$ on $\bM$, $A=1,\dots,n-1$, may be continued to a collar neighborhood of the boundary. Thus $e_{A}^j a_B^k h_{jk}=\delta_{AB}$ and $e_{Aj}e_{Bk}\delta^{AB}=h_{jk}$. The $n$th vector of the local frame on $\mathcal{M}$ is just the unit normal, $e_n=n$.

We shall still use the Greek letters $\mu,\nu,\dots$ etc and Roman letters $a,b,c$ etc to denote world and tangential indices on $\mathcal{M}$, respectively. Therefore, the formulas (\ref{Dop3}) can be used without any changes. There are several useful relations involving the extrinsic curvature $K_{ij}$ in the Gaussian coordinates
\begin{equation}
\Gamma_{jk}^n=K_{jk}=-\frac 12 \partial_n g_{jk},\qquad \Gamma_{nk}^j=-K_k^j \,.\label{Kjk}
\end{equation}
As in Sec.\ \ref{sec:3D} the semicolon will denote full covariant derivative. The colon will denote the covariant derivative with respect to Reimannian structure of $\bM$. The difference between two covariant derivatives is measured by the extrinsic curvature, e.g., $v_{j;k}=v_{j:k}-K_{jk}v_n$ for any vector $v_\mu$. Let $\tilde R_{ijkl}$ denote the Riemann tensor constructed form the boundary metric. Then the following (Gauss-Codazzi) equations are valid
\begin{eqnarray}
&&R_{ijkl}=\tilde R_{ijkl}+K_{jk}K_{il}-K_{jl}K_{ik}\,,\label{eqG}\\
&&R_{njkl}=K_{jl:k}-K_{jk:l}\,.\label{eqC}
\end{eqnarray}

\subsection{Computation of the anomaly}\label{sec:Com}
The regularity of $\eta(s,\slashed{D})$ at $s=0$ on even-dimensional manifolds with chiral bag boundary conditions (that are more general than the conditions that we consider here) was demonstrated in \cite{Gilkey:2005qm}. Let us compute the variation of $\eta(0,\slashed{D})$ under a variation of vierbein.
We consider only such variations of $e_{\mu a}$ that do not destroy the Gaussian coordinates and our assumptions on the local frame near the boundary. Namely, in the expression $\delta e_{\mu a}=u_{ab}e_{\mu b}$ only the components $u_{AB}$ will be non-zero. As in sec. \ref{sec:3D}, we assume that $u_{AB}$ is symmetric. These restrictions correspond to a partial gauge fixing of diffeomorphisms and local frame rotations. They will not affect our final result which shall be expressed through gauge invariant quantities. 

It is important to note that under general variations of the metric both boundary conditions (\ref{1stbc}) and (\ref{2ndbc}) change. Thus, in general, not only the operator $\slashed{D}$ varies in (\ref{etaint}), but also the space where we take the trace. However, under the restrictions that we have formulated in the previous paragraph, the unit normal $n$ and $\gamma^n$ remain invariant. This guarantees invariance of the first boundary condition (\ref{1stbc}). As we show in Appendix \ref{AppB} this is enough to ensure applicability of eqs.\ (\ref{deta}) -- (\ref{dWodd}). 

The variation of $\slashed{D}$ has the same functional form as in 3 dimensions consisting of a first order and a zeroth order parts, see  \eqref{varD}, (\ref{q1Q0}). One has to remember however that $u_{nn} = u_{nA} =0$. By computing the traces in (\ref{a3}) one immediately gets
\begin{equation}
a_{3}\left(Q_0,\slashed{D}^2\right) = \frac{1}{2}a_3\left(\left(\nabla\cdot q_1,\slashed{D}^2\right)\right) = 0.\label{as3D}
\end{equation}
Therefore, the only remaining contribution to the parity anomaly reads
\begin{equation}
a_{3}(\delta\slashed{D},\slashed{D}^2) = \frac{\partial }{\partial z}\big|_{z=0} a_5(1,L(z)).  \label{Final_a3}
\end{equation}
Here $L(z)\equiv L(0,z)$. 

Let us study how the invariants that enter $a_5$, see \cite{Branson:1999jz}, depend on $z$.
The connection in $L$ is varied as in Eq.\ (\ref{omz}), while the curvature of $\omega$ changes according to (\ref{Omz}). One can easily check that the nonzero components of $\mathcal{Q}_{a\mu\nu}$ are equal to:
\begin{eqnarray}
&&\mathcal{Q}_{Ajk} = \tfrac{1}{2}\left( u_{Aj:k} - u_{Ak:j} \right),\nonumber\\
&&\mathcal{Q}_{n jk} = \tfrac{1}{2}\left( u_{jB} K^B_k - u_{kB} K^B_j\right) ,\nonumber \\
&&\mathcal{Q}_{Ajn} =  -\mathcal{Q}_{Anj}= 
\tfrac{1}{2}(u_{Aj;n} - u_{AB}K^B_j ).
 \label{Qdef2}
\end{eqnarray}
Hence, 
\begin{eqnarray}
\Omega_{\mu\nu}(z) &=&  \Omega_{\mu\nu} + \ii  z  \mathcal{Q}_{a\mu\nu} \gamma^a + \mathcal{O}(z^2), \nonumber\\
\Omega_{jk:p}(z) &=& \Omega_{jk:p} +\ii  z  (\mathcal{Q}_{ajk:p}\gamma^a
 + \mathcal{Q}_{Ajk} K^A_p \gamma^n - \mathcal{Q}_{njk} K_{pB}\gamma^B  {-} \tfrac{1}{2} u_{Ap} R^{A}_{\ cjk} \gamma^c
  ) + \mathcal{O}(z^2),
\nonumber\\
 \Omega_{jn;n}(z) &=&  \Omega_{jn;n} + \ii  z  \mathcal{Q}_{ajn;n} \gamma^a + \mathcal{O}(z^2).
\end{eqnarray}

Variation of the connection $\omega$ may affect in principle all covariant derivatives of all geometric quantities. For the derivatives of $\chi$ and $S$, we have
\begin{eqnarray}
&&\chi_{:q}(z) = - \ii \varepsilon_{\alpha}\,K_{qB}\gamma^5\gamma^B, \nonumber \\
&&\chi_{:qr}(z) =  - \ii \varepsilon_{\alpha}\, K_{qB:r} \gamma^5\gamma^B - \ii\varepsilon_{\alpha}\,  K_{qB} K^{B}_r \gamma^5\gamma^n + z \varepsilon_{\alpha} u_{Ar} K^A_q \gamma^5\nonumber\\
&&S_{:q} (z)= - \tfrac{1}{4} K_{:q}  - \tfrac{\ii}{4}\,\varepsilon_{\alpha}\, K_{:q} \gamma^5\gamma^n + \tfrac{\ii}{4} \,\varepsilon_{\alpha}\, K_{qB} K\gamma^5\gamma^B, \nonumber\\
&&S_{:qk}(z)= \tfrac {\ii}4 \varepsilon_\alpha \gamma^5 \bigl( K_{:q}K_{kB}\gamma^B
+K_{:k}K_{qB}\gamma^B+K_{qB:k}K\gamma^B -K_{:qk}\gamma^n +K_{AB}K^{AB}\gamma^n\bigr)\nonumber\\
&&\qquad -\tfrac 14 K_{:qk} - \tfrac 14 z \,\varepsilon_{\alpha}\,u_{Ak} K^A_q K \gamma^5.\label{vchiS}
\end{eqnarray}
We neglected all $\mathcal{O}(z^2)$ terms. It is important to note that neither $\chi_{:q}(z)$ nor $S_{:q}(z)$ contain terms that are linear in $z$. Also, such terms do not appear in the derivatives of $E$, $K_{ij}$ and of the Riemann tensor. In other words, the terms in $a_5$ that contribute to (\ref{Final_a3}) must contain either $\Omega$ or second derivatives of $\chi$ or $S$. This reduces the number of relevant terms from about 150 \cite{Branson:1999jz} to just 21 listed in (\ref{a5}). After taking the traces, just 3 terms remain
\begin{eqnarray}
&&\frac{\partial }{\partial z}|_{z=0} \mathrm{tr}\,\left(\chi \, \Omega_{jk}\Omega^{jk}\right) 
  =     {{-\varepsilon_{\alpha}}} 2  \mathcal{Q}_{Ajk} R_{BC}^{\ \ \ jk}\epsilon^{nABC} , \nonumber\\
&&\frac{\partial }{\partial z}|_{z=0} \mathrm{tr}\,\left(\chi \, \Omega_{jn}\Omega^{jn}\right) =  
     {{-\varepsilon_{\alpha}}} 2 \mathcal{Q}_{Ajn} R_{BC}^{\ \ \ jn}\epsilon^{nABC} , \nonumber\\
&&\frac{\partial }{\partial z}|_{z=0} \mathrm{tr}\,\left(  \chi \,\chi^{:j} \chi^{:p} \,\Omega_{jp}
\right)   =  {{-\varepsilon_{\alpha}}} 4  \mathcal{Q}_{Cjp} \, K^j_A K^p_B \epsilon^{nABC} . \label{interm}
\end{eqnarray}
By combining together \eqref{Final_a3}, \eqref{Qdef2}, \eqref{interm} and \eqref{a5} and using the Gauss-Codazzi equations (\ref{eqG}), (\ref{eqC}) we obtain
\begin{eqnarray}
\delta W^{\mathrm{odd}} &=& 
-\ii \sqrt{\pi} a_3(\delta\slashed{D},\slashed{D}^2) =  
\int_{\partial\mathcal{M}} d^3x\, \sqrt{h}\, \varepsilon_{\alpha} \left\{-\frac{\ii}{192 \pi} 
u_{Aq}
  \tilde{R}_{BC~~:k}^{~~~qk}   \epsilon^{nABC} \right.
 \nonumber \\
&+& \left.\frac{\ii}{128 \pi}  
\left(
\,u_{si;n} K^i_{p:l} 
-  u_{si}  \left(K^i_l K^r_{p:r}  + K^r_p K^i_{l:r} + K^{ri} K_{rp:l}\right)
\right)
\epsilon^{nspl} \right\}. \label{result}
\end{eqnarray}

The action itself can be recovered by using (\ref{varCS}) applied to the boundary metric together with the following variational equation
\begin{eqnarray}
&&\delta\left( \int_{\partial\mathcal{M}} d^3 x \,\sqrt{h}\, K_{si}K^i_{p:l}\, \epsilon^{nspl}\right) \nonumber\\
&&\qquad = -  2\int_{\partial\mathcal{M}} d^3 x \,\sqrt{h} 
\left(
\, u_{si;n}  K^i_{p:l} 
- u_{si} \left(K^i_l K^r_{p:r}  + K^r_p K^i_{l:r} + K^{ri} K_{rp:l}\right)
\right)\nonumber
\end{eqnarray}
Hence,
\begin{equation}
W^{\mathrm{odd}}=-\frac {\ii}{4\pi} \int_{\bM} d^3x \sqrt{h} \varepsilon_\alpha 
\left[ \tfrac 1{96} \left( \widetilde \Gamma^{r}_{qi} \partial_j\widetilde \Gamma^q_{rk} + \tfrac{2}{3}\widetilde \Gamma^{r}_{qi}\widetilde \Gamma^{q}_{pj}\widetilde\Gamma^{p}_{rk} \right)\epsilon^{nijk} + \tfrac 1{64} K_{si}K^i_{p:l}\, \epsilon^{nspl}
\right] \label{final}
\end{equation}
We put tildes over Christoffel connections to stress that they have to be computed with the induced boundary metric. In this form, (\ref{final}) is valid in any coordinate system. The actions contains a Chern-Simons term for the boundary metric and a term depending on he extrinsic curvature. The properties of this action will be discussed in Sec.\ \ref{sec:prop}.

In (\ref{final}) we neglected all possible topological contributions, i.e. the terms that have vanishing local variations. To recover these terms one has to perform a direct computation of $\eta(0,\slashed{D})$ for at least one representative of a given topological class. For the ball in $\mathbb{R}^4$, for example, this has been done in \cite{Kirchberg:2006wu}.

As in 3D, one has to check that $a_j(\delta\slashed{D},\slashed{D}^2)=0$ for $j=0,1,2$ which is an easy exercise.

\subsection{Properties of the anomaly}\label{sec:prop}
\paragraph{Relation to the 3D anomaly} The 4D gravitational parity anomaly contains a boundary action only. Nevertheless, it cannot be associated to quantum effective action of a field theory on $\bM$ since it depends on $K_{ij}$ which is not intrinsic for the boundary\footnote{This does not however exclude a holographic interpretation of the 4D anomaly since the relation between asymptotic degrees of freedom and the boundary metric are more subtle.}. Even if the boundary is totally geodesic, $K_{ij}$, the 4D is not equal to the anomaly of any number of 3D fermions since the coefficient in front of Chern-Simons term in 4D, Eq.\ (\ref{final}), is $\pm 1/2$ of that in 3D. Therefore, one needs a half integer number of 3D fermions to reproduce the boundary anomaly in 4D. However, as demonstrated in \cite{Kurkov:2017cdz} at the example of fermions interacting with an abelian gauge field, this $1/2$ is just the right relation between the 3D parity anomaly and 4D boundary Chern-Simons term. The arguments of this paper are easily generalized to gravity anomalies. Indeed, let us take $\mathcal{M}=\tilde{\mathcal{M}}\times [0,\ell ]$ with $\tilde{\mathcal{M}}$ being a compact 3D manifold without boundaries. $\bM$ has two components corresponding to $x^4=0$ and $x^4=\ell$. Due to the product structure, $K_{ij}=0$. The spectrum of $\slashed{D}$ in the Kaluza-Klein limit $\ell\to 0$ may be analyzed along the lines of \cite{Kurkov:2017cdz}. There are two cases to distinguish. (i) If the sign factors $\varepsilon_\alpha$ are opposite on two components of the boundary, in the limit $\ell\to 0$ one massless 3D mode remains. The Chern-Simons terms in $W^{\rm odd}$ (\ref{final}) add up (note opposite orientations of $n$ at $x^4=0$ and $x^4=\ell$) recovering the correct coefficient\footnote{We remind here that there is a sign ambiguity in the parity anomaly related to the sign in front of $\ii\pi$ in (\ref{zetaD}). In the Pauli-Villars regularization it corresponds to the sign of mass in the factor $m/|m|$.} for the 3D parity anomaly (\ref{W3}). (ii) If both sign factors are equal, there is no massless 3D mode in the Kaluza-Klein limit, and also the Chern-Simons terms on the boundaries cancel against each other. 

\paragraph{Conformal invariance} 
Let us prove that the action (\ref{final}) is invariant under conformal (Weyl) transformations of the metric $g_{\mu\nu}\to e^{2\phi}g_{\mu\nu}$. The Chern-Simons term is known to be invariant, so that we are left with the $K$-dependent term only. For simplicity we may assume that before Weyl rescaling the metric had the Gaussian form (\ref{ds2}). The combination $\sqrt{h} \epsilon^{nijk}$ is Weyl invariant. The extrinsic curvature changes as
\begin{equation}
K_{ij}\to e^\phi (K_{ij}-h_{ij}\phi_{;n})\,.\label{WKij}
\end{equation} 
Then it is a two-line computation to check Weyl invariance of the $K$-term in (\ref{final}).
\paragraph{(No) relation to bulk topological density} Let us consider an integral over a Pontryagin type topological density
\begin{equation}
P=\tfrac 14 \int_{\mathcal{M}} d^4 x \sqrt{g}\, \epsilon^{\mu\nu\alpha\beta} R^\sigma_{\ \ \tau\mu\nu} R^\tau_{\ \ \sigma\alpha\beta}\,.\label{Pont}
\end{equation}
It can be rewritten as a boundary integral \cite{Jackiw:2003pm}
\begin{equation}
P=-\int_{\bM} d^3x \sqrt{h} \mathcal{K}^n,\label{PKn}
\end{equation}
where
\begin{eqnarray}
&&\mathcal{K}^n=\epsilon^{n\alpha\beta\gamma}\left( \Gamma_{\alpha\tau}^\sigma \partial_\beta \Gamma_{\gamma\sigma}^\tau +\tfrac 23 \Gamma_{\alpha\tau}^\sigma \Gamma_{\beta\eta}^\tau \Gamma_{\gamma\sigma}^\eta \right) \nonumber\\
&&\quad = \epsilon^{nijk}\left(\widetilde\Gamma_{il}^m\partial_j \widetilde\Gamma_{km}^l +\tfrac 23 \widetilde
\Gamma_{im}^l \widetilde\Gamma_{jp}^m \widetilde\Gamma_{kl}^p -2K_{il}K_{k:j}^l \right) \,.\label{Kn}
\end{eqnarray}
We see, that the general structure of $P$ reproduces that of the parity anomaly (\ref{final}), but the relative coefficients in front of the Chern-Simons and extrinsic curvature terms are different. Thus, regardless of the choice of the sign factors $\varepsilon_\alpha$ in boundary conditions and of a possible overall factor of $P$, the parity anomaly is not a bulk integral of the Pontryagin density.

There are many ways to generalize and extend our results.
The paper \cite{Son:2015xqa} considered generation of the Chern-Simons term in a model containing 4D gauge fields interacting with non-relativistic fermions on the boundary. Since the fermions live on the boundary  and interact with its intrinsic geometry, one-loop calculations cannot produce parity anomaly terms depending on the extrinsic curvature. Such terms, however, may be induced, at least in principle, due to higher loop effects of dynamical 4D gauge fields.

Finally we remark that Eq.\ (\ref{final}) may be considered as a parity odd boundary contribution to the Induced Gravity action. Such action has been computed in various regularizations \cite{Visser:2002ew}, including the spectral one \cite{Kurkov:2013gma}.

\acknowledgments
This work was supported by the grants 2015/05120-0 and 2016/03319-6 of the S\~ao Paulo Research Foundation (FAPESP),  by the grants 401180/2014-0 and 303807/2016-4 of CNPq, by the RFBR project 18-02-00149-A and by the Tomsk State University Competitiveness Improvement Program.

\appendix
\section{Notations, conventions and the heat kernel coefficients}\label{AppA}
Here we explain the notations and conventions that have not been defined in the main text.
We mostly use the sign conventions of \cite{Fursaev:2011zz}. For example, the Ricci tensor and the scalar curvature are defined as $R_{\mu\nu}=R^\rho_{\ \ \mu\rho\nu}$ and $R=R^\mu_{\ \ \mu}$, respectively. With our conventions the scalar curvature of unit two-sphere is $R=2$. The spin-connection reads
\begin{equation}
\sigma_{\mu ab}=\Gamma_{\mu\nu}^\rho e_{\rho a}e^\nu_b - e^\nu_b\partial_\mu e_{\nu a}\,.\label{spincon}
\end{equation}

The operator $\slashed{D}^2$ can be brought to the canonical form (\ref{Lap}) with 
\begin{equation}
\omega_\mu = \tfrac 18 \sigma_{\mu ab}[\gamma^a,\gamma^b],\qquad E=-\tfrac 14 R.\label{omE} 
\end{equation}
The curvature of $\omega$ is
\begin{equation}
\Omega_{\mu\nu}=\partial_\mu\omega_\nu - \partial_\nu\omega_\mu +[\omega_\mu,\omega_\nu]
=\tfrac 14 R_{\mu\nu ab}\gamma^a\gamma^b\,.\label{Om}
\end{equation}

Below we write relevant heat kernel coefficients for a general operator of Laplace type (\ref{Lap}) with mixed boundary conditions defined by an arbitrary $\chi$ in (\ref{chi}) and an arbitrary $S$ in the first equation of (\ref{2bc2}).

The heat kernel coefficients $a_0$ - $a_3$ for mixed boundary conditions weighted with a zeroth order operator can be found in \cite{Marachevsky:2003zb}
\begin{eqnarray}
&&a_0(Q_0,L)=\frac 1{(4\pi)^{n/2}} \int_{\mathcal{M}} d^nx \sqrt{g}\, \mathrm{tr}\, Q_0\,,\label{a0}\\
&&a_1(Q_0,L)=\frac 1{4(4\pi)^{(n-1)/2}} \int_{\bM} d^{n-1}x \sqrt{h}\, \mathrm{tr}\, (Q_0\chi)\,,\label{a1}\\
&&a_2(Q_0,L)=\frac 1{6(4\pi)^{n/2}} \left[ \int_{\mathcal{M}} d^nx \sqrt{g}\, \mathrm{tr}\, Q_0
\left( E + \tfrac 16 R\right) \right.\nonumber\\
&&\qquad \left. + \int_{\bM} d^{n-1}x \sqrt{h}\, \mathrm{tr}\, \left( 2Q_0K+12Q_0S+3Q_{0;n} \right) \right] ,\label{a2} \\
&&a_3(Q_0,L)=\frac 1{384(4\pi)^{(n-1)/2}} \int_{\bM} d^{n-1}x \sqrt{h}\, \mathrm{tr}\, \left[
Q_0\bigl( -24E +24\chi E \chi +48\chi E +48 E\chi \right. \nonumber\\
&&\qquad +16\chi R - 8\chi R_{jn}^{\ \ jn} -12\chi_{:j}\chi^{:j} +12\chi_{:j}^{\ \ j}
+192 S^2+96KS +(3+10\chi)K^2 \nonumber\\
&&\qquad \left. +(6-4\chi )K_{ij}K^{ij} \bigr) +Q_{0;n}(96S + (18-12\chi)K) +24\chi Q_{0;nn}
\right] \label{a3}
\end{eqnarray}
Here we corrected an obvious misprint\footnote{The term $192Q_{0;n}S^2$ in $a_3$ should read $Q_{0;n}K(18-12\chi)$ as in (\ref{a3}).} in \cite{Marachevsky:2003zb} that was also repeated in \cite{Kurkov:2017cdz}, but did not influence any of the results of that paper.

The coefficients $a_4$ and $a_5$ are used in the present paper only to vary them with respect to the connections according to (\ref{vara2}). Thus, it is sufficient to put $Q_0=1$ and consider connection dependent terms only. The coefficient $a_4$ for mixed boundary conditions can be found in \cite{Branson:1990xp,Vassilevich:1994we}:
\begin{eqnarray}
&&a_4(1,L)=\frac 1{360(4\pi)^{n/2}} \left[ \int_{\mathcal{M}} d^nx \sqrt{g} \mathrm{tr}\, \bigl( 60 E_{;\mu}^{\ \ \mu} +30 \Omega_{\mu\nu}\Omega^{\mu\nu}+ 12 R_{;\mu}^{\ \ \mu} \bigr) \right. \nonumber\\
&&\qquad  + \int_{\bM} d^{n-1}x \sqrt{h}\, \mathrm{tr}\, \bigl( (60+180\chi)E_{;n} +(12+30\chi)R_{;n} \nonumber\\
&&\qquad \left. + 60\chi\chi^{:j}\Omega_{jn} -12\chi_{:j}\chi^{:j}K-24\chi_{:j}\chi_{:k}K^{jk} -120 \chi_{:j}\chi^{:j}S \bigr) \right]+\dots \label{a4}
\end{eqnarray}
Ellipses denote the terms that do not depend on the connection and thus do not contribute to the anomaly.
General expression for $a_5$ is much longer. Therefore, we use more restrictive criterea (explained between eqs.\ (\ref{vchiS}) and (\ref{interm}) in the main text) to select potentially relevant terms:
\begin{eqnarray}
&&a_5(1,L) =\frac{1}{5760(4\pi)^{(n-1)/2}} \int_{\partial\mathcal{M}} d^{n-1} x\,\sqrt{h} \,\mathrm{tr}\,\left(120 \chi \, \Omega_{jk}\Omega^{jk} + 180 \chi \, \Omega_{jn}\Omega^{jn}\right.  
\nonumber\\  
&&\qquad + 30 \chi \,\chi^{:j} \chi^{:p} \,\Omega_{jp}  
+ 960 S S_{:j}^{~~j} + 240 K  S_{:j}^{~~j} + 420 K^{jk}S_{:jk} -\tfrac{105}{4} \Omega_{jk}\Omega^{jk} 
\nonumber\\
&&\qquad + \tfrac{105}{4} \chi \Omega_{jk} \chi \Omega^{jk} -45 \Omega_{jn}\Omega^{jn}
- 45 \chi \Omega_{jn} \chi \Omega^{jn} + 360 \left(\Omega^{jn}\chi S_{:j} - \Omega^{jn} S_{:j} \chi\right)\nonumber\\
&&\qquad + 45 \chi \chi_{:j}\Omega^{jn}K -180\chi_{:j}\chi_{:k}\Omega^{jk}
+ 90 \chi \chi_{:j}\Omega^{jn;n} + 120 \chi\chi_{:j}\Omega^{jk:k} + 180 \chi \chi_{:j} \Omega_{kn}K^{jk}
\nonumber\\
&&\qquad + 240 E\chi_{:j}^{~~j} - \tfrac{15}{4} \chi_{:j}^{~~j} \chi_{:k}^{~~k}
-\left. \tfrac{105}{2} \chi_{:jk}\chi^{:jk} - 15\chi_{:j}\chi^{:j}\chi_{:k}^{~~k} - \tfrac{135}{2}\chi_{:k}\chi_{:j}^{~~jk} \right)+\dots
\label{a5}
\end{eqnarray}
The coefficient $a_5$ for mixed boundary conditions was first computed in \cite{Branson:1999jz}. Ian Moss \cite{Moss:2012dp} corrected a numerical factor in front of $\chi \chi^{:j}\chi^{:p}\Omega_{jp}$. This correction is essential for our computation. We made several cross-checks that confirmed the correction by Moss.

\section{Variation of spectral functions}\label{AppB}
Let us start with some definitions. We denote $\slashed{D}_0:=\slashed{D}$ and $\slashed{D}_1:=\slashed{D}+\delta\slashed{D}$ with an infinitesimal $\delta\slashed{D}$. We assume that the 1st boundary condition (\ref{1stbc}) does not change under the variation. Let us denote by $H$ the space of smooth spinors on $\mathcal{M}$ that satisfy (\ref{1stbc}). It contains two subspaces $H_0$ and $H_1$ where also the second boundary condition (\ref{2ndbc}) is imposed with $\slashed{D}_0$ and $\slashed{D}_1$, respectively. Both $H_0$ and $H_1$ are pre-Hilbert spaces with the usual scalar products of spinors on $\mathcal{M}$ and their closures are Hilbert spaces. We shall use the same notations, $H_0$ and $H_1$, for the closures. We write the expressions appearing under the integral in the expression (\ref{etaint}) for the $\eta$ functions $\eta(s,\slashed{D}_{0,1})$ as
\begin{equation}
\mathrm{Tr}\, \left( \slashed{D}_{0,1} e^{-\tau^2\slashed{D}_{0,1}} \right)_{H_0,H_1}=
\sum_N \langle \psi^{0,1}_N,\slashed{D}_{0,1} e^{-\tau^2\slashed{D}_{0,1}} \psi^{0,1}_N \rangle_{0,1}\,,\label{B1}
\end{equation}
where $\langle\ ,\ \rangle_{0,1}$ are scalar products on $H_0$ and $H_1$, while $\psi_N^{0,1}$ denotes the elements of corresponding orthonormal bases. 

The variations of $\slashed{D}$ are induced by variations of the metric on $\mathcal{M}$. Thus, the volume element and the scalar product may change in general. Let us assume for a while that the variations are traceless, so that the scalar products on $H_0$ and $H_1$ are given by the same analytic expressions. Then the product $\langle\ ,\ \rangle_{0}$ can be used to define a Hilbert space structure on $H$. 
We write for the variation of (\ref{B1}):
\begin{eqnarray}
&&\delta \, \mathrm{Tr}\, \left( \slashed{D} e^{-\tau^2\slashed{D}} \right)=
\sum_N \langle \psi^{0}_N,\delta \left( \slashed{D} e^{-\tau^2\slashed{D}^2} \right) \psi^{0}_N \rangle_{0} \nonumber\\
&&\qquad + \sum_N \left[ \langle \delta\psi_N, \slashed{D}_0 e^{-\tau^2\slashed{D}^2_0} \psi^{0}_N \rangle_{0}  + \langle \psi^{0}_N, \slashed{D}_0 e^{-\tau^2\slashed{D}_0} \delta \psi_N \rangle_{0} \right]. \label{B2}
\end{eqnarray}
Here $\delta\psi_N=\psi_N^1-\psi_N^0$. Clearly, for a given basis $\psi_N^0$ one can always choose $\psi_N^1$ in such a way that all $\delta\psi_N$ are infinitesimally small variations.

The first sum on right hand side of (\ref{B2}) gives (\ref{etavar}) and thus leads to the desired formula (\ref{dWodd}). Let us show that the terms on the second line of (\ref{B2}) give no contribution to the variation of $\eta$ function. 

The operator $\slashed{D}$ is symmetric provided the boundary condition (\ref{1stbc}) is satisfied.
\begin{equation}
\langle \psi, \slashed{D}\psi '\rangle - \langle \slashed{D}\psi, \psi' \rangle=
-\ii \int_{\bM}d^3x \sqrt{h} \psi \gamma^n \psi' =-\ii \int_{\bM}d^3x \sqrt{h} \psi \Pi_+\gamma^n \Pi_+\psi' =0,\label{B3}
\end{equation}
where we used $\Pi_-\psi\vert_{\bM}=\Pi_-\psi'\vert_{\bM}=0$. The operator $e^{-\tau^2\slashed{D}_0^2}$ has to be viewed upon as an integral operator with a symmetric kernel 
\begin{equation*}
\sum_{\lambda_0} \psi_{\lambda_0}(x)\otimes \psi_{\lambda_0}^\dag (x') e^{-\tau^2\lambda_0^2},
\end{equation*} 
where $\{ \lambda_0,\psi_{\lambda_0} \}$ is a spectral resolution of $\slashed{D}_0$. This kernel satisfies both boundary conditions (\ref{1stbc}) and (\ref{2ndbc}) in both arguments.
Therefore, we can move $\slashed{D}_0 e^{-\tau^2\slashed{D}_0}$ in (\ref{B2}) from $\delta\psi_N$ to $\psi_N^0$. The second line of (\ref{B2}) then reads
\begin{equation}
\sum_N \left[ \langle \delta\psi_N, \slashed{D}_0 e^{-\tau^2\slashed{D}^2_0} \psi^{0}_N \rangle_{0}+ \mbox{c.\ c.} \right]. \label{B4}
\end{equation}
Next we use that $\langle \delta \psi_N,\psi_N \rangle_0 +\langle \psi_N, \delta \psi_N \rangle_0 =\delta \langle \psi_N,\psi_N \rangle_0=0$. Hence all terms under the sum in (\ref{B4}) vanish, which is the desired result. 

Let us see now how our arguments have to be changed if the variations of metric change the volume element, $\sqrt{g}\to e^{2\rho}\sqrt{g}$. Then $\langle \psi,\psi' \rangle_1=\langle e^\rho \psi,e^\rho \psi' \rangle_0$. Let us define $\slashed{D}_{1,\rho}=e^\rho \slashed{D}_1 e^{-\rho}$. It is easy to see, that $\slashed{D}_{1,\rho}$ has the same eigenvalues as $\slashed{D}_1$ and is symmetric with respect to the scalar product $\langle\ ,\ \rangle_0$. Thus we can repeat the computations presented above with $\slashed{D}_{1,\rho}$ instead of $\slashed{D}_{1}$ to obtain the formula (\ref{deta}) with 
\begin{equation}
\delta\slashed{D}=\slashed{D}_{1,\rho}-\slashed{D}_0.\label{B5}
\end{equation}
Due to the cyclicity of trace, one can omit the subscript $\rho$ in (\ref{B5}) above and use $\delta\slashed{D}=\slashed{D}_{1}-\slashed{D}_0$ in (\ref{dWodd}).

\bibliography{Gparity}
\bibliographystyle{JHEP}

\end{document}